%% file: 19asilomar_fdnn.tex
\documentclass[conference]{IEEEtran}

\usepackage[dvipsnames]{xcolor}

\usepackage{mathtools}
\usepackage{mathrsfs}
\usepackage{amssymb}
\usepackage{gensymb}
\usepackage{nicefrac}

\usepackage{siunitx}
\DeclareSIUnit{\dBm}{dBm}

\usepackage{array}
\usepackage{booktabs}
\usepackage{makecell}

\usepackage{tikz}
\usepackage{circuitikz}
\usepackage{pgfplots}
\usetikzlibrary{calc}
\usetikzlibrary{shapes}
\usetikzlibrary{matrix}
\usetikzlibrary{arrows}
\usetikzlibrary{decorations.markings}
\usetikzlibrary{spy}
\usetikzlibrary{decorations.pathmorphing}
\usetikzlibrary{decorations.markings}
\usetikzlibrary{positioning}
\usetikzlibrary{fit}
\usetikzlibrary{petri}
\usetikzlibrary{chains}
\usetikzlibrary{intersections}

\newcommand{\tikzcircle}[2][red,fill=red]{\tikz[baseline=-0.5ex]\draw[#1,radius=#2] (0,0) circle ;}%

\usepackage{pgfplots}
\pgfplotsset{compat=1.12}
\usepgfplotslibrary{units}
\usepgfplotslibrary{fillbetween}

\pgfplotsset{
    legend image with text/.style={
        legend image code/.code={%
            \node[anchor=center] at (0.3cm,0cm) {#1};
        }
    },
}

\definecolor{darkgreen}{RGB}{0,158,0}

\definecolor{Set1-7-1}{RGB}{178,24,43}
\definecolor{Set1-7-2}{RGB}{166,189,219}
\definecolor{Set1-7-3}{RGB}{214,197,115}
\definecolor{Set1-7-4}{RGB}{39,143,54}
\definecolor{Set1-7-5}{RGB}{122,1,119}
\definecolor{Set1-7-6}{RGB}{244,196,69}
\definecolor{Set1-7-7}{RGB}{0,0,0}

\usepackage{todonotes}
\presetkeys{todonotes}{inline}{}

\usepackage[noadjust]{cite}
\usepackage{threeparttable}
\usepackage{subcaption}
\usepackage{caption}

\IEEEoverridecommandlockouts

\newcommand{\andreas}[1]{\textcolor{red}{ANDREAS: #1}}
\newcommand{\andy}[1]{\textcolor{green}{ANDY: #1}}
\newcommand{\alex}[1]{\textcolor{blue}{ALEX: #1}}

\renewcommand{\andreas}[1]{}
\renewcommand{\andy}[1]{}
\renewcommand{\alex}[1]{}

\begin{document}

\bstctlcite{IEEEexample:BSTcontrol}

%
\title{Advanced Machine Learning Techniques for Self-Interference Cancellation in Full-Duplex Radios}

%
\author{\IEEEauthorblockN{Andreas Toftegaard Kristensen\IEEEauthorrefmark{1}, Andreas Burg\IEEEauthorrefmark{1}, and Alexios Balatsoukas-Stimming\IEEEauthorrefmark{2}}
\IEEEauthorblockA{\IEEEauthorrefmark{1}Telecommunication Circuits Laboratory, \'{E}cole polytechnique f\'{e}d\'{e}rale de Lausanne, Switzerland\\
\IEEEauthorrefmark{2}Department of Electrical Engineering, Eindhoven University of Technology, Netherlands}%
\thanks{This work was supported by the Swiss NSF under grant \#200021\_182621.}%
}


\maketitle

\begin{abstract}
In-band full-duplex systems allow for more efficient use of temporal and spectral resources by transmitting and receiving information at the same time and on the same frequency.
However, this creates a strong self-interference signal at the receiver, making the use of self-interference cancellation critical.
Recently, neural networks have been used to perform digital self-interference with lower computational complexity compared to a traditional polynomial model.
In this paper, we examine the use of advanced neural networks, such as recurrent and complex-valued neural networks, and we perform an in-depth network architecture exploration.
Our neural network architecture exploration reveals that complex-valued neural networks can significantly reduce both the number of floating-point operations and parameters compared to a polynomial model, whereas the real-valued networks only reduce the number of floating-point operations.
For example, at a digital self-interference cancellation of \SI{44.51}{\decibel}, a complex-valued neural network requires \SI{33.7}{\percent} fewer floating-point operations and \SI{26.9}{\percent} fewer parameters compared to the polynomial model.
\end{abstract}

\section{Introduction}\label{sec:introduction}

For beyond-5G communication systems to reach orders-of-magnitude better performance than current systems, a combination of new techniques are required.
One such technique is in-band full-duplex (FD), where information is transmitted and received simultaneously and on the same frequency band.
While FD systems have long been considered impractical due to the strong self-interference (SI) caused by the transmitter to its own receiver, more recent work on the topic (e.g.,~\cite{Jain2011,Duarte2012,Bharadia2013, Korpi2017}) has demonstrated that it is possible to achieve sufficient SI cancellation to make FD systems viable. 

The SI cancellation is usually performed in multiple steps, with the goal of reducing the SI signal to the receiver noise floor. 
The SI is usually first partially removed in the analog RF domain, applying either passive and/or active suppression to avoid saturating the analog-to-digital converter (ADC) of the receiver.
However, analog cancellation is generally expensive due to the additional analog circuitry and a residual SI signal typically still remains at the receiver, which is canceled in the digital domain. 
This requires modeling the non-linear effects of the different stages of the transceiver, such as digital-to-analog converter (DAC) and ADC non-linearities~\cite{Balatsoukas2015}, IQ imbalance~\cite{Balatsoukas2015, Korpi2014}, phase-noise~\cite{Sahai2013, Syrjala2014}, and power amplifier (PA) non-linearities~\cite{Balatsoukas2015, Korpi2014, Anttila2014, Korpi2017}.
Traditionally, this has been done using polynomial models, which have been shown to work well in practice.
However, the polynomial models have a high implementation complexity as the number of estimated parameters grows rapidly with the maximum considered non-linearity order.
As an alternative to polynomial models, neural networks (NNs) have recently been proposed for SI cancellation~\cite{Balatsoukas2018,kurzo2018design} where it was shown that NNs can achieve similar SI cancellation performance with a polynomial model with significantly lower computational complexity.

\subsubsection*{Contribution}

In this work, we revisit and extend the work of~\cite{Balatsoukas2018}. 
More specifically, in addition to real-valued (feed-forward) NNs (RVNNs), we also consider recurrent neural networks (RNNs) and complex-valued neural networks (CVNNs). 
Moreover, we perform an in-depth network architecture exploration to evaluate the performance of the different NNs as a function of the number of floating-point operations (FLOPs) and the number of NN parameters.
This exploration shows that the CVNNs consistently require fewer parameters than their real-valued counterparts for the same SI cancellation performance, indicating that the complex-valued representation is more powerful for the SI cancellation problem.
Moreover, we also show that the CVNNs can reduce both the number of FLOPs and parameters compared to a polynomial model, whereas the various real-valued networks (RVNNs) typically only reduce the number of floating-point operations and increase the number of parameters.
For example, at an SI cancellation of \SI{44.45}{\decibel}, a CVNN requires \SI{33.7}{\percent} fewer FLOPs and \SI{26.9}{\percent} fewer parameters than the polynomial model, while an RVNN (specifically, a real-valued RNN) requires \SI{30.5}{\percent} fewer FLOPs but \SI{69.2}{\percent} more parameters. 

\begin{figure}[t]
	\centering
	\scalebox{0.9}{\input{fig/system.tikz}}
	\caption{Simplified wireless FD transceiver block diagram~\cite{kurzo2018design}.}
	\label{fig:block}
\end{figure}
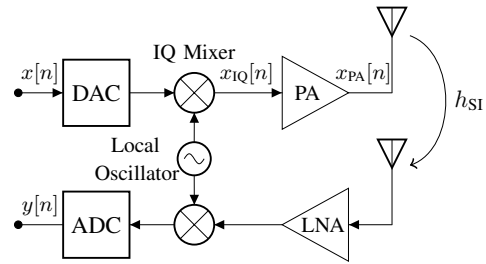

\section{Background}\label{sec:background}

In this section, we first describe the main task of digital SI cancellation.
Then, we provide an overview of the existing methods for digital SI cancellation and we provide some background on RNNs and CVNNs.

\subsection{Self-Interference Cancellation}

A basic block diagram of an FD transceiver is shown in Fig.~\ref{fig:block}.
If a signal-of-interest from a remote node is not present and assuming that some form of analog SI cancellation is performed, the received signal $y[n]$ is the SI signal that needs to be canceled in the digital domain.
The goal of the digital SI cancellation is to reconstruct an estimate of the SI signal $y[n]$, denoted $\hat{y}[n]$, which is a non-linear function of the transmitted baseband samples.  
The estimated SI signal $\hat{y}[n]$ is then subtracted from $y[n]$ so that the residual SI signal is $e[n] = y[n] - \hat{y}[n]$.
Then, the amount of SI cancellation for a window of length $N$, expressed in dB, is given by $C_{\text{dB}} = 10\log_{10}\left( \frac{\sum _{n=0}^{N-1}|y[n]|^2}{\sum _{n=0}^{N-1}|e[n]|^2}\right)$.

\subsubsection{Non-Linear Polynomial SI Cancellation}

For the polynomial SI canceller, the dominant non-idealities to consider are the IQ-mixer and the power amplifier (PA) in the transmitter~\cite{Anttila2014,Korpi2017}, while the remaining transceiver components are assumed to be ideal.
With these assumptions, a state-of-the-art polynomial SI cancellation model is given as~\cite{Korpi2017}
\begin{align}
	y[n]	& = \sum _{\substack{p=1,\\p \text{ odd}}}^P \sum_{q=0}^p\sum_{l=0}^{L-1}h_{p,q}[l] x[n-l]^{q}x^*[n-l]^{p-q} , \label{eq:final}
\end{align}
where $x[n]$ is the transmitted digital baseband signal, $L$ corresponds to the overall memory of the system, and $P$ is the non-linearity order.
The $\hat{h}_{p,q}$ parameters contain the combined effects from the IQ-mixer, the PA, and the SI channel and they can be obtained using least-squares estimation.

\begin{figure}[t!]
  \centering
  \scalebox{0.8}{\input{fig/nn.tikz}}
  \caption{Example of an FFNN for the reconstruction of the non-linear component of the SI signal~\cite{Balatsoukas2018}. The network is composed of input- (\tikzcircle[black!20!green, fill=black!20!green]{3pt}), hidden- (\tikzcircle[blue!75, fill=blue!75]{3pt}), and output-nodes (\tikzcircle[red!75, fill=red!75]{3pt}).}\label{fig:nn}
\end{figure}
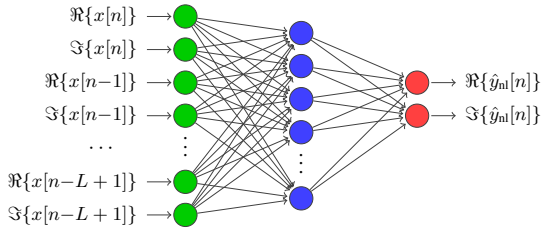

\subsubsection{Feed-Forward Neural Network SI Cancellation}\label{sec:ffnn}

As shown in Fig.~\ref{fig:nn}, a feed-forward NN (FFNN) has three types of nodes: input nodes, hidden nodes, and output nodes.
The input to each node of a given layer is a weighted sum of the outputs of the nodes in the previous layer, and the output of each node is obtained by applying a non-linear \emph{activation} function to this weighted sum. 
The SI signal can be conceptually decomposed as $y[n] = y_{\text{lin}}[n] + y_{\text{nl}}[n]$, where $y_{\text{lin}}[n]$ represents the linear part and $y_{\text{nl}}[n]$ the non-linear part.
The linear part can be estimated using standard linear cancellation, 
while the remaining part, ${y}_{\text{nl}}$, is estimated using the NNs and the two terms are added to re-generate the SI cancellation signal $\hat{y}[n] = \hat{y}_{\text{lin}}[n] + \hat{y}_{\text{nl}}[n]$. 
Since the baseband samples are complex-valued, the real and imaginary parts are split and provided as separate inputs to all the RVNNs as shown in Fig.~\ref{fig:nn}. 

\subsection{Advanced Neural Networks}
\subsubsection{Recurrent Neural Networks}

The transceiver chain shown in Fig.~\ref{fig:block} have memory, motivating the use of recurrent neural networks (RNNs), a different type of NN architecture more suitable for learning from sequential data.
For a fully-connected RNN with a single recurrent layer and one output layer, the computation for each time-step is given by
\begin{align}
	\mathbf{h}[n] &= \sigma\left( \mathbf{W}_h \mathbf{x}[n] + \mathbf{U}_h \mathbf{h}[n-1] + \mathbf{b}_h \right) , \label{eq:rnn_hidden} \\
	\mathbf{y}[n] &= \mathbf{W}_y \mathbf{h}[n] + \mathbf{b}_y , \label{eq:rnn_output}
\end{align}
where $\mathbf{h}[n]$ is the RNN state at time $n$ with input $\mathbf{x}[n]$ and $\mathbf{y}[n]$ is the RNN output. 
The activation function in the recurrent layer is typically a hyperbolic tangent, $\sigma(x) = \frac{e^{2x}-1}{e^{2x}+1}$. 
For the RNN, $L$ time-steps of inputs are used to generate the final state $\mathbf{h}[L-1]$, which is then provided to the output layer.

\subsubsection{Complex-Valued Neural Networks}

While complex-valued signals can be represented by considering the real and imaginary parts separately, this representation generally does not respect the phase information that is captured by complex algebra.
This motivates the use of complex-valued neural networks (CVNNs), where both the network parameters and operations are complex-valued.
However, a real-valued scalar loss function of complex variables is not complex-analytic and therefore not complex-differentiable, thus raising the question of how a CVNN can be trained. 
One solution is to use Wirtinger calculus~\cite{wirtinger1927formalen} (or $\mathbb{C}\mathbb{R}$ calculus~\cite{kreutz2009complex}) to obtain the complex-valued gradients required to train a CVNN.

Let $z\in \mathbb{C}$ and $f(z) \in \mathbb{R}$. The direction of steepest ascent for $f$ is then given as the derivative with respect to $z^*$, i.e., $\frac{\partial f(z)}{\partial z^*}$~\cite{bouboulis2010wirtinger}.
Using Wirtinger calculus, the derivative $\frac{\partial f(z)}{\partial z^*}$ can be calculated by re-writing $f(z)$ as a bi-variate function of $z$ and $z^*$, $f(z, z^*)$, and then treat $z^*$ as the variable and $z$ a constant.
Alternatively, $\mathbb{C}$ can be regarded as $\mathbb{R}^2$, and the complex derivatives can be obtained by considering the partial derivatives with respect to the real and imaginary parts
\begin{equation}
	\frac{\partial f}{\partial z} \triangleq \frac{1}{2} \left( \frac{\partial f}{\partial x} - j \frac{\partial f}{\partial y}  \right) \  \mbox{and} \  \frac{\partial f}{\partial z^*} \triangleq  \frac{1}{2} \left( \frac{\partial f}{\partial x} + j \frac{\partial f}{\partial y}  \right) \label{eq:wirtr2}
\end{equation}
Numerous activation functions have been proposed in the literature for CVNNs.
Table~\ref{tab:activations} shows the different activation functions considered in this work.

\begin{table}[t!]
	\centering
	\caption{CVNN activation functions.}
	\label{tab:activations}
	\begin{tabular}{ll}
		\toprule
		 Amp-Phase~\cite{hirose2012complex}      	& $f(z) = \mathrm{tanh}(|z|) \exp(i \theta_z)$ \\[0.5em]
		 Cardioid~\cite{virtue2017better}    		& $f(z) = \frac{1}{2} (1+\cos(\theta_z)) z $ \\[0.5em]
		 modReLU~\cite{arjovsky2016unitary}      	& $f(z) = \max(0, |z| + b ) \exp(i \theta_z)$ \\[0.5em]
		 $\mathbb{C}$ReLU~\cite{trabelsi2017deep}   & $f(z) = \text{ReLU}(\Re(z)) + i \text{ReLU}(\Im(z))$ \\[0.5em]
		 $z$ReLU~\cite{hirose2012complex}        	& $ f(z) = \begin{cases} z & \theta_z \in [0, \pi / 2] \\ 0 \end{cases}$\\
		 \bottomrule
	\end{tabular}
\end{table}


%

\section{Network Architecture Exploration Methodology}\label{sec:methods}

In this section, we briefly describe the data set and our methodology for the NN architecture exploration.

\subsection{Data Set}

The data set consists of QPSK-modulated OFDM signals with a pass-band bandwidth of \SI{10}{\mega\hertz} and $N_c=1024$ carriers, sampled at \SI{20}{\mega\hertz}.
Each transmitted OFDM frame consists of $\sim$\SI{20\,000}{} baseband samples, with \SI{90}{\percent} used for training and the remaining \SI{10}{\percent} used for testing. 
An average transmit power of \SI{10}{\dBm} is used and the employed two-antenna setup provides a passive analog suppression of \SI{53}{\decibel}.
Active analog cancellation is not used as the achieved passive suppression is sufficient for this work.
The test-bed and the data set are described in more detail in~\cite{Balatsoukas2015} and \cite{Balatsoukas2018}, respectively.

\begin{figure*}[t!]
	\centering
	\begin{subfigure}[t]{.34\textwidth}
		\centering
		\scalebox{1.0}{\input{fig/train_plot_nn.tikz}}
		\caption{Training of shallow and deep FFNNs.}
		\label{fig:train_ffnn}
	\end{subfigure}%
	\begin{subfigure}[t]{.32\textwidth}
		\centering
		\scalebox{1.0}{\input{fig/train_plot_cvnn.tikz}}
		\caption{Training of shallow and deep CVNNs.}
		\label{fig:train_cvnn}
	\end{subfigure}%
	\hspace*{0.06cm}%
	\begin{subfigure}[t]{.32\textwidth}
		\centering
		\scalebox{1.0}{\input{fig/train_plot_rnn.tikz}}
		\caption{Training of shallow and deep RNNs.}
		\label{fig:train_rnn}
	\end{subfigure}%
\caption{Mean cancellation on the non-linear test set for 20 different random network initializations $\pm 1$ standard deviation.}
\label{fig:train_nl}
\end{figure*}
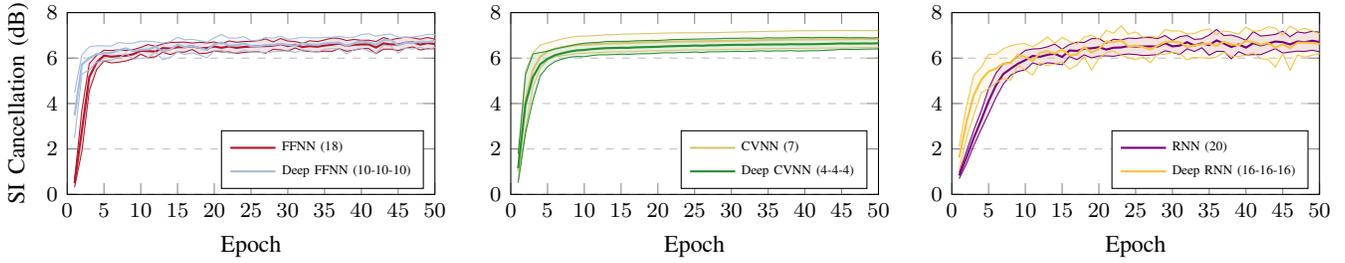

\subsection{Experimental Setup}

For the NN experiments, the various NNs are considered for different widths and depths to determine their performance as a function of their computational complexity and memory complexity.
We use $W$ to denote the width (i.e., the number of neurons) of a layer. The notation $W{-}W{-}W$ is used to indicate an NN with 3 hidden layers and $W$ neurons in each layer.
All NNs also have an output layer, with the RVNNs having 2 outputs for the real and imaginary parts and the CVNNs having only 1 complex-valued output.
For the FFNNs and RNNs, we consider shallow networks of widths $W=2, 4, \hdots, 20$, and for the deep FFNNs and RNNs we consider the widths $W{-}W{-}W = 2{-}2{-}2, 4{-}4{-}4, \hdots, 20{-}20{-}20$.
For the CVNNs, we consider the sizes $W=1, 2, \hdots, 10,$ and for the deep CVNNs we consider the sizes $W{-}W{-}W = 1{-}1{-}1, 2{-}2{-}2, \hdots, 10{-}10{-}10$.
The polynomial model in~\eqref{eq:final} is considered for powers $P=3, 5, 7, 9$.
For all NNs and the polynomial models, we use $L=13$ as in the previous work on NNs for SI cancellation with the same data-set~\cite{Balatsoukas2018}.
For each of the considered NNs, a hyperparameter search is performed using a grid-search for each NN to select the best values of the learning rate and the batch size.
A total of 20 points are sampled from the hyperparameter space, and 5-fold cross-validation is used on the training set for the hyperparameter search with 5 random weight initializations per fold.
For the learning-rate, we sample from a continuous uniform distribution $\mathrm{unif}(10^{-6}, 0.05)$ and for the batch-size, we sample from a discrete uniform distribution $\mathrm{unif}(4, 64)$.
The NNs are then trained on the entire training set with the best hyperparameters and with 20 different random initializations to see observe the effect of the initialization on the performance.
All models are implemented using Tensorflow, which has built-in support for complex-valued operations. 
All models use the Adam optimizer~\cite{kingma2014adam}, with the default values for all parameters except for the batch-size and learning-rate, which are selected during the hyperparameter search.

\begin{table}[t!]
	\centering
	\caption{Mean non-linear cancellation on the test set $\pm$1 standard deviation for the activation functions using a CVNN with 1 hidden layer of 10 units.}
	\label{tab:results_activations}
	\begin{tabular}{lllll}
		\toprule
		Amp-Phase 		& Cardioid 				& modReLU 		& $\mathbb{C}$ReLU 		& $z$ReLU \\
		\midrule
		$7.0\pm0.3$ 	& $\mathbf{7.8\pm0.1}$ 	& $3.1\pm0.3$ 	& $7.5\pm0.2$ 	& $1.8\pm0.4$ \\
	    \bottomrule
	\end{tabular}
\end{table}

\section{Results}\label{sec:results}

In this section, we present the results of the network architecture exploration. Specifically, we first briefly discuss the results of training CVNNs with different activation functions, we then consider the training behavior of the FFNNs, RNNs, and CVNNs and, finally, we present the SI cancellation performance of the different NNs as a function of the number of FLOPs and parameters.

\subsection{Complex-Valued Activation Functions}

Table~\ref{tab:results_activations} shows the results of training the same CVNN with 1 hidden layer of $W=10$ neurons for the different activation functions given in Table~\ref{tab:activations}.
The best performance is obtained with the cardioid and $\mathbb{C}$ReLU activation functions.
However, the cardioid function is only marginally better than the $\mathbb{C}$ReLU function, while being more expensive to compute.
The modReLU and $z$ReLU functions have the worst performance, similarly to~\cite{trabelsi2017deep} where the $\mathbb{C}$ReLU function in all cases performed significantly better than the modReLU and $z$ReLU functions.
The Amp-Phase activation function also achieves a relatively high performance level, however, we observed during training that the convergence to a good performance level is much slower than the cardioid and $\mathbb{C}$ReLU functions.
The cardioid and $\mathbb{C}$ReLU functions reach \SI{6}{\decibel} of non-linear cancellation within the first 5 epochs, with the Amp-Phase function requiring 10 epochs.
Therefore, in the remainder of this section, we use the $\mathbb{C}$ReLU function for all CVNNs.

\subsection{Training}

\begin{figure}[t!]
	\centering
	\begin{subfigure}[t]{\columnwidth}
		\centering
		\scalebox{1.0}{\input{fig/train_plot_nn_all.tikz}}
		\caption{Training of shallow and deep FFNNs.}
		\label{fig:train_plot_nn_all}
	\end{subfigure}
	
	\begin{subfigure}[t]{\columnwidth}
		\centering
		\scalebox{1.0}{\input{fig/train_plot_cvnn_all.tikz}}
		\caption{Training of shallow and deep CVNNs.}
		\label{fig:train_plot_cvnn_all}
	\end{subfigure}
	\caption{Mean cancellation on the test set without linear cancellation for 20 different random network initializations $\pm 1$ standard deviation.}
	\label{fig:train_all}
\end{figure}
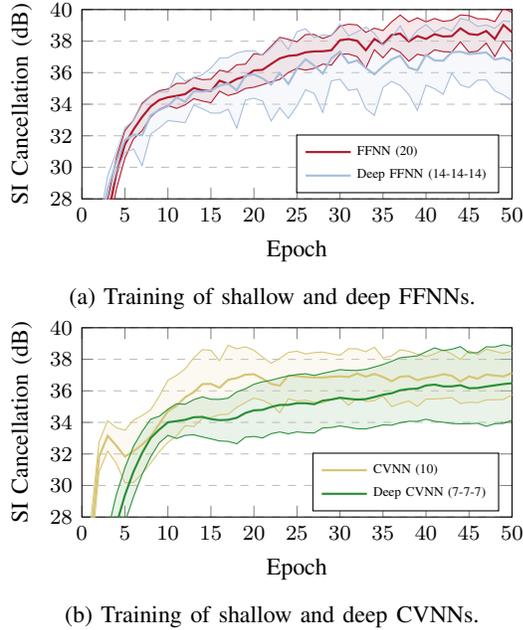

In principle, the NNs could learn to provide the full (i.e., linear and non-linear) cancellation, but, as we show in this section, this is difficult in practice.
Fig.~\ref{fig:train_all} shows the results of training some shallow and deep FFNNs and CVNNs without first applying a linear SI canceler.
The SI linear canceler alone achieves \SI{37.9}{\decibel} cancellation and the NNs are unable to achieve a performance much higher than this, even when using deeper networks.
Essentially, $y_{\text{nl}}$ is significantly weaker than $y_{\text{lin}}$, so the noise in the gradient computation hides the non-linear structure from the learning algorithm.

Fig.~\ref{fig:train_nl} shows the performance for the NNs when the linear cancellation $y_{\text{lin}}$ is performed separately from the NNs, and the NNs instead learn to cancel only $y_{\text{nl}}$.
In this case, the total cancellation achieved is much higher than what is possible when the NNs have to learn to perform the full cancellation.
This shows that for the problem of SI cancellation, the inclusion of expert knowledge is essential. 
We observe that the FFNNs in Fig.~\ref{fig:train_ffnn} and the CVNNs in Fig.~\ref{fig:train_cvnn} achieve their peak performance after a small number of epochs and remain relatively stable, whereas the RNNs in Fig.~\ref{fig:train_rnn} require more epochs to converge and show a much higher variation, both across different initializations and from one epoch to the next. 

\subsection{Complexity vs Cancellation Performance}

\begin{figure}[t!]
\centering
\begin{subfigure}{\columnwidth}
	\centering
	\scalebox{1.0}{\input{fig/SI_modern_nns.tikz}}
	\caption{Performance as a function of \# real-valued FLOPs.}
	\label{fig:perf_flops}
\end{subfigure}%

\begin{subfigure}{\columnwidth}
	\centering
	\scalebox{1.0}{\input{fig/mem_modern_nns.tikz}}
	\caption{Performance as a function of \# real-valued parameters.}
	\label{fig:perf_params}
\end{subfigure}
\caption{Test set performance for all models as a function of their computational and memory complexity.}
\label{fig:perf_compare}
\end{figure}
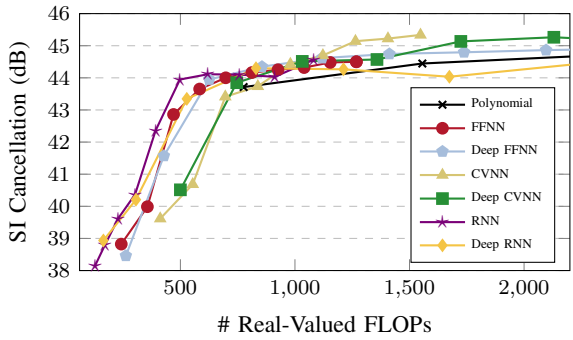
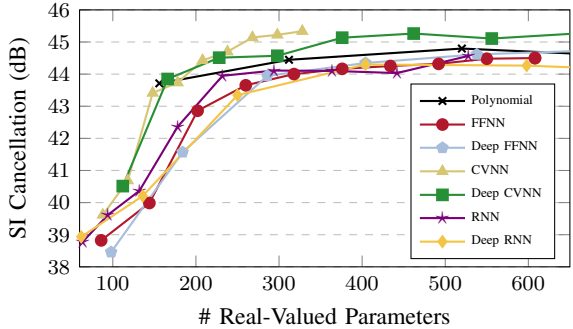

In Fig.~\ref{fig:perf_compare}, we present the test set performance of the different NNs trained with the hyperparameters selected from the search.
Fig.~\ref{fig:perf_flops} and Fig.~\ref{fig:perf_params} show the total (i.e., linear and non-linear) SI cancellation performance with respect to the number of real-valued FLOPs and the number of real-valued parameters, respectively. 
The number of real-valued FLOPs and the number of real-valued parameters are a proxy for the computational complexity and the memory requirement of the various SI cancelers, respectively. 
The number of FLOPs is the sum of the number of equivalent real-valued additions, multiplications, and applications of activation functions.
Complex-valued additions and multiplications are converted to real-valued operations assuming that one complex multiplication can be implemented using 3 real multiplications and 5 real additions and one complex addition can be implemented using 2 real additions.
For the NN-based cancelers, the number of FLOPs and parameters include the linear canceler.
Moreover, to perform a best-case complexity analysis for the polynomial canceler, it is assumed that the calculation of the polynomial basis functions in~\eqref{eq:final} comes at no computational cost.
Finally, we only count the \emph{use} of an activation function, as the specifics of how exactly an activation function is computed is an implementation detail that is difficult to abstract otherwise.

For the performance as a function of the number of FLOPs, we observe that all the models cluster closely together at an SI cancellation around \SI{44}{\decibel}.
However, it is clear the CVNNs require much fewer parameters than the other models for the same performance.
Finally, the RNNs do not provide significant performance improvements compared to the other models and there is no clear benefit of using deeper NNs.

The polynomial model performance peaks around \SI{44.8}{\decibel} for $P=9$, but this requires a very large number of FLOPs that puts it outside the figure.
For the sake of comparing a specific performance point, we consider the polynomial model for $P=5$ that achieves an SI cancellation of \SI{44.45}{\decibel}, and then we take the closest NNs which are either equal to or better than the polynomial model in performance. 
The improvement of the various NN-based cancelers over the polynomial-based canceler is shown in Table~\ref{tab:comparison}.
We observe that the FFNNs reduce the number of FLOPs, but increase the number of parameters significantly.
Moreover, the CVNNs are the only NNs capable of reducing both the number of FLOPs and the number of parameters.
In general, we can observe from Fig.~\ref{fig:perf_compare} that for SI cancellation higher than \SI{43.5}{\decibel} only the CVNNs remain better than the polynomial model for both the number of FLOPs and the number of parameters, whereas the other NNs generally use more parameters than the polynomial model and only reduce the number of FLOPs.

\subsection{Maximum Cancellation Performance}

In Fig.~\ref{fig:perf_compare}, we observe that the CVNNs achieve the highest cancellation performance of all the considered SI cancellers. 
This indicates that for the problem of SI cancellation, the CVNNs seem to have more representational power. 
However, the RVNNs still achieve a maximum performance close to that of the CVNNs.
As shown earlier, one of the best activation functions for the CVNNs is the $\mathbb{C}$ReLU function, which applies the ReLU separately to the real and imaginary parts.
It is expected that this type of activation function works well for problems where some symmetry or a special meaning on the real and imaginary parts is required~\cite{hirose2012complex}.
It is also implied in~\cite{hirose2012complex} that a network processing $n$-dimensional information with activation functions similar to $\mathbb{C}$ReLU, has neural dynamics similar to that of an RVNN processing $2n$-dimensional real-valued information, where the real and imaginary parts are dealt with separately and independently.
This may provide a more intuitive explanation as to why the maximum performance of the CVNNs and RVNNs is similar.

\section{Conclusion}\label{sec:conclusion}

In this paper, we provided an investigation into the trade-offs of applying different NNs at different performance goals for SI cancellation in FD radios.
We showed that CVNNs consistently require much fewer parameters than their real-valued equivalents for the same SI cancellation performance, indicating that the complex-valued representation is more powerful for this problem.
Moreover, we showed that only CVNNs can reduce both the number of parameters and FLOPs compared to the polynomial SI canceller.
The use of the $\mathbb{C}$ReLU activation function for the CVNNs also indicates that the neural dynamics of the CVNNs may resemble that of the RVNNs, providing an intuitive explanation as to why the difference in maximum performance is relatively close for the CVNN and RVNNs. 
Overall, when performing a careful NN architecture exploration combined with hyperparameter tuning, the results are somewhat less favorable (while still being very good in certain cases) than our initial results reported in~\cite{Balatsoukas2018}, where we only considered a single NN architecture with fixed hyperparameters and a single SI cancellation performance point. This highlights the importance of a rigorous comparison that is unfortunately often overlooked in the literature that applies NNs to communications problems.

\begin{table}[t!]
	\centering
	\caption{Reduction of \# FLOPs and \# parameters relative to the polynomial model at an SI cancellation of \SI{44.45}{\decibel}. The polynomial model has 1556 FLOPs and 312 parameters.}
	\label{tab:comparison}
	\begin{tabular}{lccc}
	\toprule	
	~ 						& Cancellation 			& \# FLOPs 	    		& \# Parameters \\
	\midrule
    FFNN (18)        		& \SI{44.48}{\decibel} 	& {$-27.5\,\%$}   		& {$+76.3\%$} \\
    Deep FFNN (10-10-10)   	& \SI{44.61}{\decibel}	& {${-29.8\%}$}  		& {$+72.4\%$} \\
    CVNN (7)       			& \SI{44.69}{\decibel} 	& {$-27.9\%$}     		& {${-23.7\%}$} \\
    Deep CVNN (4-4-4)   	& \SI{44.51}{\decibel}	& {$\mathbf{-33.7\%}$}  & {$\mathbf{-26.9\%}$} \\
    RNN (20)       			& \SI{44.57}{\decibel}	& {$-30.5\%$}  			& {$+69.2\%$} \\
    Deep RNN (16-16-16)    	& \SI{44.49}{\decibel}	& {$+82.4\%$} 			& {$+355.1\%$} \\    
    \bottomrule
	\end{tabular}
\end{table}


\bibliographystyle{IEEEtran}
\bibliography{IEEEabrv,bibliography}


\end{document}

%% file: fig/system.tikz
\begin{circuitikz}[scale=0.65]

	




	\draw (4,0) node[mixer,scale=0.6] (txmixer) {};
	\draw (8.5,0) node[antenna,scale=0.6] (txantenna) {};
	
	\draw (0,0) to[short,*-] ++(0,0) to[twoport,>,t=DAC] (txmixer.west) node[inputarrow]{};
	\draw (txmixer.east) to[short,-] ++(1.55,0) to[amp,>,t=\small{PA}] ++(1.5,0) to[short,-] (txantenna);
	
	\draw (0.5,0) node[above] {\small $x[n]$};
	\draw (txmixer)+(0,0.5) node[above] {\small{IQ Mixer}};
	\draw (txmixer)+(1.2,0) node[above] {\small $x_{\text{IQ}}[n]$};
	\draw (txantenna)+(-0.65,0) node[above] {\small $x_{\text{PA}}[n]$};
	
	\draw (8.5,-3) node[antenna,scale=0.6] (rxantenna) {};
	\draw (4,-3) node[mixer,scale=0.6] (rxmixer) {};
	
	\draw (rxantenna) to[short,-] ++(-1.,0) to[amp,>,t={\rotatebox[origin=c]{180}{\small{LNA}}}] ++(-1.5,0) to[short,-] (rxmixer.east) node[inputarrow,rotate=180]{};
	\draw (rxmixer.west) to[twoport,>,t=ADC] (0,-3) to[short,-*] (0,-3);
	\draw [->] (txantenna)+(0.4,1.3) to[thick, out=-20, in=20, edge node={node [right] {$h_{\text{SI}}$}}]  ($(rxantenna) + (0.4,1.35)$);
	
	\draw (0.5,-3) node[above] {\small $y[n]$};
	
	\draw (4.375,-1.5) node[oscillator,scale=0.5] (ref) {};
	\draw (ref.south) to[short,-] (rxmixer.north) node[inputarrow,rotate=270]{};
	\draw (ref.north) to[short,-] (txmixer.south) node[inputarrow,rotate=90]{};
	\draw (ref)+(-0.6,0) node[left,text width=1.2cm,align=center] {\small{Local\\Oscillator}};
	
\end{circuitikz}

%% file: fig/nn.tikz
\def\layersep{3.5cm}
\def\twolayersep{7cm}
\def\inputnodes{6}
\def\hiddennodes{5}
\def\outputnodes{2}

\begin{tikzpicture}[shorten >=1pt,->,draw=black!75, node distance=\layersep, scale=0.55]
    \tikzstyle{every pin edge}=[<-,shorten <=1pt]
    \tikzstyle{neuron}=[circle,draw=black!75,fill=black!75,minimum size=11pt,inner sep=0pt]
    \tikzstyle{input neuron}=[neuron, fill=black!20!green];
    \tikzstyle{output neuron}=[neuron, fill=red!75];
    \tikzstyle{hidden neuron}=[neuron, fill=blue!75];
    \tikzstyle{annot} = [text width=10em, text centered]

	\node[input neuron, pin=left: {\small$\Re{\left\{x[n]\right\}}$}] (I-1) at (0,-1) {};
	\node[input neuron, pin=left: {\small$\Im{\left\{x[n]\right\}}$}] (I-2) at (0,-2) {};
	\node[input neuron, pin=left: {\small$\Re{\left\{x[n{-}1]\right\}}$}] (I-3) at (0,-3) {};
	\node[input neuron, pin=left: {\small$\Im{\left\{x[n{-}1]\right\}}$}] (I-4) at (0,-4) {};
	\node at (-2.5,-5) {$\cdots$};
	\node at (0,-4.75) {$\vdots$};
	\node[input neuron, pin=left: {\small$\Re{\left\{x[n{-}L+1]\right\}}$}] (I-5) at (0,-6) {};
	\node[input neuron, pin=left: {\small$\Im{\left\{x[n{-}L+1]\right\}}$}] (I-6) at (0,-7) {};

	\pgfmathsetmacro{\limit}{\hiddennodes-1}
	\foreach \name / \y in {1,...,\limit}
		\path[yshift=-0.5cm]
			node[hidden neuron] (H-\name) at (\layersep,-\y cm) {};
	\node at (\layersep,-5.25) {$\vdots$};
	\path[yshift=-0.5cm]
		node[hidden neuron] (H-\hiddennodes) at (\layersep,-6 cm) {};

	\foreach \name / \y in {1,...,\outputnodes}
		\path[yshift=-2cm]
			node[output neuron,pin={[pin edge={->}]right:\ifodd\y {\small $\Re{\left\{\hat{y}_{\text{nl}}[n]\right\}}$} \else {\small $\Im{\left\{\hat{y}_{\text{nl}}[n]\right\}}$} \fi}] (O-\name) at (\twolayersep,-\y) {};

    \foreach \source in {1,...,\inputnodes}
        \foreach \dest in {1,...,\hiddennodes}
			\path (I-\source) edge (H-\dest);

    \foreach \source in {1,...,\hiddennodes}
		\foreach \dest in {1,...,\outputnodes}
			\path (H-\source) edge (O-\dest);

\end{tikzpicture}

%% file: fig/train_plot_nn.tikz
\begin{tikzpicture}
    \begin{axis}[
		normalsize,
		width = 6.47cm,
		height = 4cm,
		xmin=0, xmax=50,
		ymin=0, ymax=8,
		ymajorgrids=true,
		grid style=dashed,
		xlabel = {Epoch},
		ylabel = {SI Cancellation (dB)},
		ylabel near ticks,
		xlabel near ticks,
		xtick distance=5,
		ytick distance=2,
		label style={font=\small},
		tick label style={font=\footnotesize},
		ymajorgrids,
		legend pos = south east,
		legend style={font=\tiny},
		legend columns=1,
		legend cell align=left,
    ]

	\addplot[Set1-7-1, thick, solid] table[x index = 0, y index = 1] {fig/results_adam/pgf_dat_train/ffnn_nl_struct_18_test_cancellation_n_epochs_50_train_size_1_0_batch_size_39_lr_0_003887654113612277.dat};
    \addlegendentry{FFNN (18)}
	\addplot+[name path=ffnnbot, color=Set1-7-1, mark=none, forget plot] table[x index = 0, y index = 2] {fig/results_adam/pgf_dat_train/ffnn_nl_struct_18_test_cancellation_n_epochs_50_train_size_1_0_batch_size_39_lr_0_003887654113612277.dat};
	\addplot+[name path=ffnntop, color=Set1-7-1, mark=none, forget plot] table[x index = 0, y index = 3] {fig/results_adam/pgf_dat_train/ffnn_nl_struct_18_test_cancellation_n_epochs_50_train_size_1_0_batch_size_39_lr_0_003887654113612277.dat};
	\addplot[Set1-7-1!50,fill opacity=0.2, forget plot] fill between[of=ffnnbot and ffnntop];

	\addplot[Set1-7-2, thick, solid] table[x index = 0, y index = 1] {fig/results_adam/pgf_dat_train/ffnn_nl_struct_10-10-10_test_cancellation_n_epochs_50_train_size_1_0_batch_size_38_lr_0_004859727512044656.dat};
 \addlegendentry{Deep FFNN (10-10-10)}
	\addplot+[name path=dffnnbot, color=Set1-7-2, mark=none, forget plot] table[x index = 0, y index = 2] {fig/results_adam/pgf_dat_train/ffnn_nl_struct_10-10-10_test_cancellation_n_epochs_50_train_size_1_0_batch_size_38_lr_0_004859727512044656.dat};
	\addplot+[name path=dffnntop, color=Set1-7-2, mark=none, forget plot] table[x index = 0, y index = 3] {fig/results_adam/pgf_dat_train/ffnn_nl_struct_10-10-10_test_cancellation_n_epochs_50_train_size_1_0_batch_size_38_lr_0_004859727512044656.dat};
	\addplot[Set1-7-2!50,fill opacity=0.2, forget plot] fill between[of=dffnnbot and dffnntop];

    \end{axis}
\end{tikzpicture}

%% file: fig/train_plot_cvnn.tikz
\begin{tikzpicture}
    \begin{axis}[
		normalsize,
		width = 6.47cm,
		height = 4cm,
		xmin=0, xmax=50,
		ymin=0, ymax=8,
		ymajorgrids=true,
		grid style=dashed,
		xlabel = {Epoch},
		ylabel near ticks,
		xlabel near ticks,
		xtick distance=5,
		ytick distance=2,
		label style={font=\small},
		tick label style={font=\footnotesize},
		ymajorgrids,
		legend pos = south east,
		legend style={font=\tiny},
		legend columns=1,
		legend cell align=left,
    ]

	\addplot[Set1-7-3, thick, solid] table[x index = 0, y index = 1] {fig/results_adam/pgf_dat_train/complex_ffnn_nl_struct_7_test_cancellation_n_epochs_50_train_size_1_0_batch_size_60_lr_0_009062472075216842.dat};
    \addlegendentry{CVNN (7)}
	\addplot+[name path=cvnnbot, color=Set1-7-3, mark=none, forget plot] table[x index = 0, y index = 2] {fig/results_adam/pgf_dat_train/complex_ffnn_nl_struct_7_test_cancellation_n_epochs_50_train_size_1_0_batch_size_60_lr_0_009062472075216842.dat};
	\addplot+[name path=cvnntop, color=Set1-7-3, mark=none, forget plot] table[x index = 0, y index = 3] {fig/results_adam/pgf_dat_train/complex_ffnn_nl_struct_7_test_cancellation_n_epochs_50_train_size_1_0_batch_size_60_lr_0_009062472075216842.dat};
	\addplot[Set1-7-3!50,fill opacity=0.2, forget plot] fill between[of=cvnnbot and cvnntop];

	\addplot[Set1-7-4, thick, solid] table[x index = 0, y index = 1] {fig/results_adam/pgf_dat_train/complex_ffnn_nl_struct_4-4-4_test_cancellation_n_epochs_50_train_size_1_0_batch_size_39_lr_0_003887654113612277.dat};
	\addlegendentry{Deep CVNN (4-4-4)}
	\addplot+[name path=dcvnnbot, color=Set1-7-4, mark=none, forget plot] table[x index = 0, y index = 2] {fig/results_adam/pgf_dat_train/complex_ffnn_nl_struct_4-4-4_test_cancellation_n_epochs_50_train_size_1_0_batch_size_39_lr_0_003887654113612277.dat};
	\addplot+[name path=dcvnntop, color=Set1-7-4, mark=none, forget plot] table[x index = 0, y index = 3] {fig/results_adam/pgf_dat_train/complex_ffnn_nl_struct_4-4-4_test_cancellation_n_epochs_50_train_size_1_0_batch_size_39_lr_0_003887654113612277.dat};
	\addplot[Set1-7-4!50,fill opacity=0.2, forget plot] fill between[of=dcvnnbot and dcvnntop];

    \end{axis}
\end{tikzpicture}

%% file: fig/train_plot_rnn.tikz
\begin{tikzpicture}
    \begin{axis}[
		normalsize,
		width = 6.47cm,
		height = 4cm,
		xmin=0, xmax=50,
		ymin=0, ymax=8,
		ymajorgrids=true,
		grid style=dashed,
		xlabel = {Epoch},
		ylabel near ticks,
		xlabel near ticks,
		xtick distance=5,
		ytick distance=2,
		label style={font=\small},
		tick label style={font=\footnotesize},
		ymajorgrids,
		legend pos = south east,
		legend style={font=\tiny},
		legend columns=1,
		legend cell align=left,
    ]

	\addplot[Set1-7-5, thick, solid] table[x index = 0, y index = 1] {fig/results_adam/pgf_dat_train/rnn_simple_nl_struct_20_test_cancellation_n_epochs_50_train_size_1_0_batch_size_158_lr_0_0023234741855871666.dat};
    \addlegendentry{RNN (20)}
	\addplot+[name path=cvnnbot, color=Set1-7-5, mark=none, forget plot] table[x index = 0, y index = 2] {fig/results_adam/pgf_dat_train/rnn_simple_nl_struct_20_test_cancellation_n_epochs_50_train_size_1_0_batch_size_158_lr_0_0023234741855871666.dat};
	\addplot+[name path=cvnntop, color=Set1-7-5, mark=none, forget plot] table[x index = 0, y index = 3] {fig/results_adam/pgf_dat_train/rnn_simple_nl_struct_20_test_cancellation_n_epochs_50_train_size_1_0_batch_size_158_lr_0_0023234741855871666.dat};
	\addplot[Set1-7-5!50,fill opacity=0.2, forget plot] fill between[of=cvnnbot and cvnntop];

	\addplot[Set1-7-6, thick, solid] table[x index = 0, y index = 1] {fig/results_adam/pgf_dat_train/rnn_simple_nl_struct_16-16-16_test_cancellation_n_epochs_50_train_size_1_0_batch_size_152_lr_0_004859727512044656.dat};
	\addlegendentry{Deep RNN (16-16-16)}
	\addplot+[name path=dcvnnbot, color=Set1-7-6, mark=none, forget plot] table[x index = 0, y index = 2] {fig/results_adam/pgf_dat_train/rnn_simple_nl_struct_16-16-16_test_cancellation_n_epochs_50_train_size_1_0_batch_size_152_lr_0_004859727512044656.dat};
	\addplot+[name path=dcvnntop, color=Set1-7-6, mark=none, forget plot] table[x index = 0, y index = 3] {fig/results_adam/pgf_dat_train/rnn_simple_nl_struct_16-16-16_test_cancellation_n_epochs_50_train_size_1_0_batch_size_152_lr_0_004859727512044656.dat};
	\addplot[Set1-7-6!50,fill opacity=0.2, forget plot] fill between[of=dcvnnbot and dcvnntop];

    \end{axis}
\end{tikzpicture}

%% file: fig/train_plot_nn_all.tikz
\begin{tikzpicture}
    \begin{axis}[
		normalsize,
		width = 7.3cm,
		height = 4.1cm,
		xmin=0, xmax=50,
		ymin=28, ymax=40,
		ymajorgrids=true,
		grid style=dashed,
		xlabel = {Epoch},
		ylabel = {SI Cancellation (dB)},
		ylabel near ticks,
		xlabel near ticks,
		xtick distance=5,
		ytick distance=2,
		label style={font=\small},
		tick label style={font=\footnotesize},
		ymajorgrids,
		legend pos = south east,
		legend style={font=\tiny},
		legend columns=1,
		legend cell align=left,
    ]

	\addplot[Set1-7-1, thick, solid] table[x index = 0, y index = 1] {fig/results_adam/pgf_dat_train/ffnn_all_struct_20_test_cancellation_n_epochs_50_train_size_1_0_batch_size_39_lr_0_0023234741855871666.dat};
    \addlegendentry{FFNN (20)}
	\addplot+[name path=ffnnbot, color=Set1-7-1, mark=none, forget plot] table[x index = 0, y index = 2] {fig/results_adam/pgf_dat_train/ffnn_all_struct_20_test_cancellation_n_epochs_50_train_size_1_0_batch_size_39_lr_0_0023234741855871666.dat};
	\addplot+[name path=ffnntop, color=Set1-7-1, mark=none, forget plot] table[x index = 0, y index = 3] {fig/results_adam/pgf_dat_train/ffnn_all_struct_20_test_cancellation_n_epochs_50_train_size_1_0_batch_size_39_lr_0_0023234741855871666.dat};
	\addplot[Set1-7-1!50,fill opacity=0.2, forget plot] fill between[of=ffnnbot and ffnntop];

	\addplot[Set1-7-2, thick, solid] table[x index = 0, y index = 1] {fig/results_adam/pgf_dat_train/ffnn_all_struct_14-14-14_test_cancellation_n_epochs_50_train_size_1_0_batch_size_39_lr_0_0023234741855871666.dat};
 \addlegendentry{Deep FFNN (14-14-14)}
	\addplot+[name path=dffnnbot, color=Set1-7-2, mark=none, forget plot] table[x index = 0, y index = 2] {fig/results_adam/pgf_dat_train/ffnn_all_struct_14-14-14_test_cancellation_n_epochs_50_train_size_1_0_batch_size_39_lr_0_0023234741855871666.dat};
	\addplot+[name path=dffnntop, color=Set1-7-2, mark=none, forget plot] table[x index = 0, y index = 3] {fig/results_adam/pgf_dat_train/ffnn_all_struct_14-14-14_test_cancellation_n_epochs_50_train_size_1_0_batch_size_39_lr_0_0023234741855871666.dat};
	\addplot[Set1-7-2!50,fill opacity=0.2, forget plot] fill between[of=dffnnbot and dffnntop];

    \end{axis}
\end{tikzpicture}

%% file: fig/train_plot_cvnn_all.tikz
\begin{tikzpicture}
    \begin{axis}[
		normalsize,
		width = 7.3cm,
		height = 4.1cm,
		xmin=0, xmax=50,
		ymin=28, ymax=40,
		ymajorgrids=true,
		grid style=dashed,
		xlabel = {Epoch},
		ylabel = {SI Cancellation (dB)},
		ylabel near ticks,
		xlabel near ticks,
		xtick distance=5,
		ytick distance=2,
		label style={font=\small},
		tick label style={font=\footnotesize},
		ymajorgrids,
		legend pos = south east,
		legend style={font=\tiny},
		legend columns=1,
		legend cell align=left,
    ]

	\addplot[Set1-7-3, thick, solid] table[x index = 0, y index = 1] {fig/results_adam/pgf_dat_train/complex_ffnn_all_struct_10_test_cancellation_n_epochs_50_train_size_1_0_batch_size_22_lr_0_004884508028205187.dat};
    \addlegendentry{CVNN (10)}
	\addplot+[name path=cvnnbot, color=Set1-7-3, mark=none, forget plot] table[x index = 0, y index = 2] {fig/results_adam/pgf_dat_train/complex_ffnn_all_struct_10_test_cancellation_n_epochs_50_train_size_1_0_batch_size_22_lr_0_004884508028205187.dat};
	\addplot+[name path=cvnntop, color=Set1-7-3, mark=none, forget plot] table[x index = 0, y index = 3] {fig/results_adam/pgf_dat_train/complex_ffnn_all_struct_10_test_cancellation_n_epochs_50_train_size_1_0_batch_size_22_lr_0_004884508028205187.dat};
	\addplot[Set1-7-3!50,fill opacity=0.2, forget plot] fill between[of=cvnnbot and cvnntop];

	\addplot[Set1-7-4, thick, solid] table[x index = 0, y index = 1] {fig/results_adam/pgf_dat_train/complex_ffnn_all_struct_7-7-7_test_cancellation_n_epochs_50_train_size_1_0_batch_size_39_lr_0_0023234741855871666.dat};
	\addlegendentry{Deep CVNN (7-7-7)}
	\addplot+[name path=dcvnnbot, color=Set1-7-4, mark=none, forget plot] table[x index = 0, y index = 2] {fig/results_adam/pgf_dat_train/complex_ffnn_all_struct_7-7-7_test_cancellation_n_epochs_50_train_size_1_0_batch_size_39_lr_0_0023234741855871666.dat};
	\addplot+[name path=dcvnntop, color=Set1-7-4, mark=none, forget plot] table[x index = 0, y index = 3] {fig/results_adam/pgf_dat_train/complex_ffnn_all_struct_7-7-7_test_cancellation_n_epochs_50_train_size_1_0_batch_size_39_lr_0_0023234741855871666.dat};
	\addplot[Set1-7-4!50,fill opacity=0.2, forget plot] fill between[of=dcvnnbot and dcvnntop];
    \end{axis}
\end{tikzpicture}

%% file: fig/SI_modern_nns.tikz
\begin{tikzpicture}

	\pgfplotsset{grid style={dashed}}

	\begin{axis}[
		normalsize,
		width = 8.1cm,
		height = 5cm,
		xlabel = {\# Real-Valued FLOPs},
		ylabel = {SI Cancellation (dB)},
		ylabel near ticks,
		xlabel near ticks,
		ytick distance=1,
		label style={font=\small},
		tick label style={font=\footnotesize},
		yticklabel style={
						/pgf/number format/fixed,
						/pgf/number format/precision=0,
						/pgf/number format/zerofill
		},
		scaled y ticks=false,
		xmin = 60, xmax = 2200,
		ymin = 38, ymax = 46,
		ymajorgrids,
		legend pos = south east,
		legend style={font=\tiny},
		legend columns=1,
		legend cell align=left,
		legend entries={Polynomial, FFNN, Deep FFNN, CVNN, Deep CVNN, RNN, Deep RNN},
	]
	\addplot[black, thick, solid, mark=x] table[x index = 0, y index = 1] {fig/results_poly/pgf_dat_complexity/flop_polynomial_nl_complexity_mult_algo_reduced_cmult.dat};
	\addplot[Set1-7-1, thick, solid, mark=*] table[x index = 0, y index = 1] {fig/results_adam/pgf_dat_complexity/flop_ffnn_nl_complexity_mult_algo_reduced_cmult_shallow.dat};
	\addplot[Set1-7-2, thick, solid, mark=pentagon*] table[x index = 0, y index = 1] {fig/results_adam/pgf_dat_complexity/flop_ffnn_nl_complexity_mult_algo_reduced_cmult_deep.dat};
	\addplot[Set1-7-3, thick, solid, mark=triangle*] table[x index = 0, y index = 1] {fig/results_adam/pgf_dat_complexity/flop_complex_ffnn_nl_complexity_mult_algo_reduced_cmult_shallow.dat};
	\addplot[Set1-7-4, thick, solid, mark=square*] table[x index = 0, y index = 1] {fig/results_adam/pgf_dat_complexity/flop_complex_ffnn_nl_complexity_mult_algo_reduced_cmult_deep.dat};
	\addplot[Set1-7-5, thick, solid, mark=text, text mark=$\star$] table[x index = 0, y index = 1] {fig/results_adam/pgf_dat_complexity/flop_rnn_simple_nl_complexity_mult_algo_reduced_cmult_shallow.dat};
	\addplot[Set1-7-6, thick, solid, mark=diamond*] table[x index = 0, y index = 1] {fig/results_adam/pgf_dat_complexity/flop_rnn_simple_nl_complexity_mult_algo_reduced_cmult_deep.dat};
	\end{axis}
\end{tikzpicture}

%% file: fig/mem_modern_nns.tikz
\begin{tikzpicture}

	\pgfplotsset{grid style={dashed}}

	\begin{axis}[
		normalsize,
		width = 8.1cm,
		height = 5cm,
		xlabel = {\# Real-Valued Parameters},
		ylabel = {SI Cancellation (dB)},
		ylabel near ticks,
		xlabel near ticks,
		ytick distance=1,
		label style={font=\small},
		tick label style={font=\footnotesize},
		yticklabel style={
						/pgf/number format/fixed,
						/pgf/number format/precision=0,
						/pgf/number format/zerofill
		},
		scaled y ticks=false,
		xmin = 60, xmax = 650,
		ymin = 38, ymax = 46,
		ymajorgrids,
		legend pos = south east,
		legend style={font=\tiny},
		legend columns=1,
		legend cell align=left,
		legend entries={Polynomial, FFNN, Deep FFNN, CVNN, Deep CVNN, RNN, Deep RNN}
	]

	\addplot[black, thick, solid, mark=x] table[x index = 0, y index = 1] {fig/results_poly/pgf_dat_complexity/mem_polynomial_nl_complexity_mult_algo_reduced_cmult.dat};
	\addplot[Set1-7-1, thick, solid, mark=*] table[x index = 0, y index = 1] {fig/results_adam/pgf_dat_complexity/mem_ffnn_nl_complexity_mult_algo_reduced_cmult_shallow.dat};
	\addplot[Set1-7-2, thick, solid, mark=pentagon*] table[x index = 0, y index = 1] {fig/results_adam/pgf_dat_complexity/mem_ffnn_nl_complexity_mult_algo_reduced_cmult_deep.dat};
	\addplot[Set1-7-3, thick, solid, mark=triangle*] table[x index = 0, y index = 1] {fig/results_adam/pgf_dat_complexity/mem_complex_ffnn_nl_complexity_mult_algo_reduced_cmult_shallow.dat};
	\addplot[Set1-7-4, thick, solid, mark=square*] table[x index = 0, y index = 1] {fig/results_adam/pgf_dat_complexity/mem_complex_ffnn_nl_complexity_mult_algo_reduced_cmult_deep.dat};
	\addplot[Set1-7-5, thick, solid, mark=text, text mark=$\star$] table[x index = 0, y index = 1] {fig/results_adam/pgf_dat_complexity/mem_rnn_simple_nl_complexity_mult_algo_reduced_cmult_shallow.dat};
	\addplot[Set1-7-6, thick, solid, mark=diamond*] table[x index = 0, y index = 1] {fig/results_adam/pgf_dat_complexity/mem_rnn_simple_nl_complexity_mult_algo_reduced_cmult_deep.dat};
	\end{axis}

\end{tikzpicture}